\documentclass[10pt,a4paper,twocolumn]{article}
\usepackage[utf8]{inputenc}
\usepackage{amsmath}
\usepackage{amsthm}
\numberwithin{equation}{section}
\usepackage{amsfonts}
\usepackage{mathtools}
\usepackage{mathrsfs}
\usepackage{mathpazo}
\usepackage{multirow}
\usepackage{amssymb}
\usepackage{graphicx}
\usepackage{ifpdf}
\usepackage{braket}
\usepackage[dvipsnames]{xcolor}

\usepackage{cite}
\usepackage[bookmarks=true,colorlinks=true,linkcolor=orange,citecolor=orange,urlcolor=orange,bookmarksnumbered]{hyperref}

\usepackage[left=1.5cm, right=1.5cm, top=2cm, bottom=3cm]{geometry}
\begin{document}
\twocolumn[
 \begin{@twocolumnfalse}

\begin{center}

\vspace{0.5truecm}

{\Large \bf The Manakov-Zakharov-Ward model \\
as an integrable decoupling limit of the membrane}

\vspace{0.5truecm}

{David Osten}

\vspace{0.3truecm}

{\em Institute for Theoretical Physics (IFT), University of Wroc\l aw \\
pl. Maxa Borna 9, 50-204 Wroc\l aw, Poland
}

\vspace{0.2truecm}

{{\tt david.osten@uwr.edu.pl}}

\vspace{0.2truecm}
\end{center}

\begin{abstract}
\noindent A novel decoupling limit of the membrane is proposed, leading to the $(1+2)$-dimensional classically integrable model originally introduced by Manakov, Zakharov, and Ward. This limit is the large-wrapping regime of a membrane propagating toy background of the form $\mathbb{R}_t \times T^2 \times G$ subject to scaling limit, where $G$ is a Lie group and the geometry is supported by a four-form flux. Such toy backgrounds can arise from consistent eleven-dimensional supergravity solutions, exemplified by the uplift of the pure NSNS AdS$_3 \times$ S$^3 \times$ T$^4$ background. The scaling limit can be interpreted as similtaneous small tension and non- or hyper-relativistic limit.
\end{abstract}
\vspace{0.5truecm}

 \end{@twocolumnfalse}
]

\section{Introduction}
M-theory and its relation to the membrane world-volume theory remain enigmatic except in a few special limits. In the infinite‑momentum frame, the BFSS matrix model describes eleven‑dimensional dynamics as the quantum mechanics of \(N\) D0‑branes, yet it lacks manifest Lorentz invariance and requires \(N\to\infty\) to recover the full continuum theory \cite{Banks:1996vh}. Its low‑energy two‑derivative approximation is captured by eleven‑dimensional supergravity \cite{Bergshoeff:1987cm}. In the weak‑coupling regime, a wrapped membrane on a spatial circle yields the perturbative Type IIA string $\sigma$-model \cite{Duff:1987bx}. The ABJM-construction \cite{Aharony:2008ug}, the holographic dual to M-theory in AdS\(_4\times \)S\(^7\), provides a method for this particular background, but does not directly probe generic eleven-dimensional geometries. Recently, new relations between the (super)membrane world-volume theory and quantities in the ABJM-theory have been obtained in \cite{Giombi:2023vzu,Beccaria:2023ujc} and follow-up papers by these authors. 

In addition to these previously mentioned understood corners of membrane theory, this letter suggests \textit{a novel tractable regime} -- via a decoupling limit of the membrane in a certain background geometry -- described by a $(1+2)$-dimensional classically integrable theory. Integrable field theories in $(1+1)$ dimensions evade typical no-go constraints by admitting infinitely many local conserved charges that render scattering purely elastic and factorised. However, in $(1+2)$ or higher dimensions, the same infinite symmetries force the S-matrix to be trivial (that is, free), as any non-trivial scattering would violate the independence from impact parameters. Thus, beyond one spatial dimension, integrable models must break Lorentz invariance. One of the few existing examples of such non-Lorentz invariant integrable theories is the Manakov-Zakharov-Ward model \cite{Manakov:1981,Ward:1988}.

A non-Lorentzian decoupling limit, as proposed here, fits into a recently observed pattern that certain decoupling limits that isolate BPS \(p\)‑brane dynamics and associated matrix models \cite{Andringa:2012uz,Harmark:2018cdl,Bergshoeff:2018yvt}, naturally giving rise to non‑Lorentzian target‑space geometries known as \(p\)-brane Newton-Cartan geometry (see \cite{Oling:2022} for a recent review). More recently, this has also been discussed extended to eleven-dimensional supergravity \cite{Lambert:2024yjk,Lambert:2024ncn,Blair:2023noj,Blair:2024aqz}.

\subsection{The bosonic membrane}
The action of the bosonic membrane (or equivalently the bosonic part of the supermembrane  \cite{Bergshoeff:1987cm}) in a bosonic background of 11d supergravity, described by a metric $g$ and three-form gauge field $C$ is
\begin{align}
    S &= T \int \mathrm{d}^3 \sigma \left( \sqrt{- \gamma} + C_3 \right) \label{eq:MembraneAction}
\end{align}
in its Nambu-Goto type form with $\gamma = \text{det}(\gamma_{MN})$ and
\begin{align*}
  \gamma_{MN} &= \partial_M \hat{X}^\mu \partial_N \hat{X}^\nu g_{\mu \nu} , \\
   C_3 &=  \frac{1}{3!} \epsilon^{KLM} \partial_K \hat{X}^\mu \partial_L \hat{X}^\nu \partial_M \hat{X}^\rho C_{\mu \nu \rho}.
\end{align*}
$\sigma^M$ are the coordinates on the membrane world-volume, $\hat{X}^\mu(\sigma)$ the embedding coordinate fields, and $T$ the membrane tension. 

Due to the square-root, the equations of motion are highly non-linear and quite intractable. Even a Polyakov-like formulation, presented in appendix \ref{app:Polyakov}, is not advantageous here as the world-volume metric cannot be gauge-fixed to conformally flat. Famously \cite{deWit:1988wri,deWit:1988xki}, even this supposedly most simple case leads to complications, such as a continuous spectrum and spike instabilities. Quantum non-integrability of the membrane in the sense of scattering of fluctuations around a classical solution, and its relation to string theory, has recently been shown in \cite{Seibold:2024oyr}. In this letter, we will see that a large wrapping limit together with a scaling limit of the background can simplify these equations. In the special case considered here, the result will even be a \textit{classically integrable system}.

\subsection{The Manakov-Zakharov-Ward model}
Classically integrable field theories in dimensions higher than $(1+1)$ are rare, especially those of the $\sigma$-model type, which are abundant in $(1+1)$ dimensions (see \cite{Hoare:2021dix,Demulder:2023bux} for recent reviews). A notable exception is the Manakov-Zakharov-Ward (MZW) model \cite{Manakov:1981,Ward:1988}, an integrable field theory in $(1+2)$ dimensions. It arises via dimensional reduction from the anti-self-dual Yang-Mills equations in $(2+2)$-dimensional spacetime. This reduction involves assuming that the gauge potentials are independent of one spatial coordinate, resulting in a \textit{non-Lorentz invariant} system in $(1+2)$ dimensions.\footnote{As a recent approach to obtain new integrable models in higher dimensions, let us mention \cite{Sergyeyev_2025}.}

The model is of $\sigma$-model type and can be formulated in terms of maps $g: \Sigma \rightarrow G$ from a $(1+2)$-dimensional flat spacetime $\Sigma$ with standard coordinates $\sigma^M$ (referred to as the world-volume) into a Lie group $G$. The dynamics is conveniently expressed using the left-invariant Maurer–Cartan current $j = g^{-1} \mathrm{d}g \in \mathfrak{g}$, valued in the Lie algebra $\mathfrak{g}$ of $G$. The flatness condition $\mathrm{d}j + \frac{1}{2}[j, j] = 0$ is automatically satisfied. The equations of motion take the form
\begin{equation}
    \mathrm{d} \star j + \mathrm{d}j \wedge v = \mathrm{d} \star j - \frac{1}{2} [j, j] \wedge v = 0, \label{eq:MZW}
\end{equation}
where $v = \mathrm{d}\sigma^y$ -- with $\sigma^y$ denoting a distinguished coordinate for $y= 0,1$ or $2$ -- is a constant world-volume one-form with unit norm. The appearance of $v$ in the equations of motion manifestly breaks Lorentz invariance of this model in $(1+2)$ dimensions. Let us also present an action that gives rise to the equations of motion \eqref{eq:MZW}
\begin{equation}
    S_{MZW} = T_0 \int \left( \frac{1}{2} \text{tr}(j \wedge \star j) + \frac{1}{3} \sigma^y \ \text{tr}( j \wedge [j \overset{\wedge}{,} j] ) \right)
\end{equation}
where tr denotes some ad-invariant non-generate bilinear form on $\mathfrak{g}$. To the knowledge of the author, this action has not been presented in the literature before.

Originally introduced by Manakov and Zakharov \cite{Manakov:1981} for the case of $G=\mathrm{SU}(2)$ and a $v$ being timelike, $v = \mathrm{d}\sigma^0$, the model was shown to be solvable by the inverse scattering transform. Ward \cite{Ward:1988} demonstrated that for the model to have a well-defined energy functional, the vector $v$ must be \textit{space-like}, although its integrability remains unaffected by the nature of $v$. Depending on the nature of $v$ the model remains with $(1+1)$-dimensional Lorentz or two-dimensional Euclidean symmetry. Choosing $v=\mathrm{d}\sigma^2$ and coordinates\footnote{with light-cone coordinates $\sigma^\pm = \frac{1}{2} ( \sigma^0 \pm \sigma^1)$} $(\sigma^+,\sigma^-,\sigma^y$), the equations of motion can be written explicitly as
\begin{equation}
    \partial_- j_+ = \partial_y j^y .
\end{equation}
These equations can be obtained as the integrability condition of the auxiliary linear system \cite{Ward:1988}
\begin{align}
    A(\sigma) \tilde{g}(\sigma,z) &= (z \partial_y - \partial_+) \tilde{g}(\sigma,z) \\
    B(\sigma) \tilde{g}(\sigma,z) &= (z \partial_- - \partial_y) \tilde{g}(\sigma,z)
\end{align}
with a complex spectral parameter $z$ and appropriate reality conditions. The group valued field $g$ is reproduced as $g(\sigma) = \tilde{g}(\sigma,0)$. Using this overdetermined linear system, one can construct soliton solutions. The MZW model supports soliton solutions localised in all spatial directions and does not show classical scattering for lump solitons, an identifying feature of integrability \cite{Ward:1988}. Ioannidou and Ward constructed an infinite sequence of non-local but finite conserved charges using integrals of the currents and its powers \cite{Ioannidou:1995he}. These charges reflect the underlying loop-algebra and twistor-theoretic structure of the model \cite{Mason:1990}, and parallel the hierarchy of integrable flows found in systems such as the matrix KP-hierarchy \cite{Dimakis:2007wv}. This integrable structure also governs soliton interactions and scattering phenomena \cite{Ioannidou:1999}.

Recently, the MZW-model has reemerged in the context of higher-dimensional Chern–Simons theory \cite{Costello:2017dso,Costello:2018gyb}, a unifying framework for integrable $\sigma$-models associated with Lie group symmetries. In this setting, the MZW-model appears as a special case in certain higher-dimensional generalisations \cite{Bittleston:2020hfv,Schenkel:2024dcd}. Partly because of its breaking of Lorentz invariance, it was not obvious in which physical settings the MZW-model appears. Moreover, it seemed unlikely that the MZW-model will play a role in membrane dynamics since it assumes a flat metric on $\Sigma$ in contrast to the dynamical induced metric of the bosonic membrane. However, this article will show that the MZW-model can be obtained as a large-wrapping decoupling limit of the bosonic membrane.

\section{A decoupling limit for the membrane on T$^2$}
In the limit where a bosonic membrane wraps a two-torus \( \omega \) times with \( \omega \rightarrow \infty \), the effective world-volume coupling scales as \( 1/\omega \), rendering the theory asymptotically free. This behaviour has been observed in different settings: for example, in de Wit-Hoppe–Nicolai matrix regularisation where the membrane is described by an SU(\( N \)) matrix model with coupling \( g \propto T \sim 1/\omega \) \cite{deWit:1988ig}, or the stretched wrapped membrane setting in \cite{Bjornsson:2004yp,Bjornsson:2005sh}. Let us present an incarnation of this setting in the classical membrane world-volume theory.

For this, let us consider the bosonic membrane in a toy model background -- in the sense, that for now we do not focus on whether this is a solution to eleven-dimensional supergravity:
\begin{equation}
    \mathcal{M}_\tau \times \mathcal{M}_H = (\mathbb{R}_t \times T^2) \times \mathcal{M}_H \label{eq:toymodelgeometry}
\end{equation}
supported by a certain background 4-form flux. We make the following ansatz for the metric and three-form gauge field, introducing scales $R$ and $\bar{R}$ for the longitudinal ($\mathcal{M}_\tau$, coordinates\footnote{In a slight abuse of notation, we use the same indices $K,L,M,...$ for the transversal coordinates $\hat{X}^M$ and the world-volume coordinates $\sigma^M$.} $\hat{X}^M$) and transversal spaces ($\mathcal{M}_H$, coordinates $\hat{X}^m$)
\begin{align}
    \mathrm{d}s^2 &= R^2 \eta_{MN} \mathrm{d}\hat{X}^M \mathrm{d}\hat{X}^N + \bar{R}^2 H_{mn}(\hat{X}^k) \ \mathrm{d}\hat{X}^m \mathrm{d}\hat{X}^n , \label{eq:ToyModelBackgroundWrapping} \\
    F &= \frac{1}{3!} R \bar{R}^2 F_{klmN}(\hat{X}^n) \mathrm{d}\hat{X}^k \wedge \mathrm{d}\hat{X}^l \wedge \mathrm{d}\hat{X}^m \wedge \mathrm{d} \hat{X}^N . \nonumber
\end{align}
for a flat metric $\eta$ and a -- for the purposes of this section -- unspecified metric $H$ on $\mathcal{M}_H$. The scaling of $F$ has to be introduced in order to be consistent with the supergravity equations -- compare \eqref{eq:AdS3S3uplift} later. Moreover, we note the particular form of $F$ with one leg in $\mathcal{M}_\tau$ and the other in $\mathcal{M}_H$. In this section, we consider \eqref{eq:ToyModelBackgroundWrapping} as a toy model without considering it as a solution to the supergravity equations of motion.

On the world-volume, we assume arbitrary (finite) fluctuations around a \textit{static wrapped membrane} in a large wrapping number limit $\omega \rightarrow \infty$:
\begin{align}
\hat{X}^\mu(\sigma) &= \left( \hat{X}^M(\sigma) , \hat{X}^m(\sigma) \right) \label{eq:ExpansionWrapping} \\
&= \left( \omega \sigma^M + X^M(\sigma) , X^m(\sigma) \right). \nonumber
\end{align}
The parameter $\omega$ describes both a 'world-volume speed of light' and the wrapping number around the torus.\footnote{In principle, these parameters could be different, of course -- i.e. $ X^M = ( \omega_0 \sigma^0, \omega_1 \sigma^1 , \omega_2 \sigma^2)$. For the limit in question, we want to assume $\omega_0 \sim \omega_1 \sim \omega_2 \sim \omega \rightarrow \infty$. Finite quotients could be compensated for by rescaling of the membrane coordinates $\sigma^M$.} In particular, we aim at the following, \textit{at leading order} in this limit:
\begin{itemize}
    \item decoupling limit of the (internal three-dimensional gravity of the) membrane, i.e $\gamma_{MN} \rightarrow \eta_{MN}$
    \item membrane dynamics along $\mathcal{M}_\tau$ and $\mathcal{M}_H$ decouple
\end{itemize}
This will involve scaling limits -- both on the target space and on the membrane world-volume. In section \ref{chap:MZWfromMembrane} it is shown which concrete choice of $\mathcal{M}_H$ and $F$ leads to the MZW-model dynamics \eqref{eq:MZW} and how this can be embedded into a supergravity solution.

\paragraph{The membrane action in the large wrapping limit.} 
In the limit $\omega \rightarrow \infty$, the $\mathcal{O}(\omega^0)$ world-volume fields $X^\mu(\sigma)$ are suppressed, and the action \eqref{eq:MembraneAction} simplifies significantly. In particular, 
\begin{align}
    \gamma_{MN} &= \omega^2 R^2 \left( \eta_{MN} + \frac{1}{\omega} 2 \partial_{(M} X_{N)}   \right. \label{eq:InducedMetricExpansion} \\
    &{} \quad \left. + \frac{1}{\omega^2} \partial_M X^K \partial_N X^L \eta_{KL}+\frac{\bar{R}^2}{R^2 \omega^2} \partial_M X^k \partial_N X^l H_{kl} \right), \nonumber  \\
    \sqrt{-\gamma} &= \omega^3 R^3 \left( 1 + \frac{1}{\omega} \partial_M X^M + \right. \label{eq:InducedMetricDetExpansion} \\
    &{} \quad + \frac{1}{\omega^2} \left( \partial_K X_L \partial^K X^L + (\partial_K X^K)^2 \right)  \nonumber \\
    &{} \quad + \left. \frac{\bar{R}^2}{R^2 \omega^2} \frac{1}{2} \eta^{MN} \partial_M X^k \partial_N X^l H_{kl} + \mathcal{O}\left( \frac{1}{\omega^3},\frac{\bar{R}^4}{R^4 \omega^4} \right) \right) \nonumber.
\end{align}
For this expansion to make sense, we also have to assume the following scaling:
\begin{equation}
    \frac{\bar{R}^2}{R^2 \omega^2} \rightarrow 0.
\end{equation}
For writing the membrane action in this background, it is advantageous that $F$ in \eqref{eq:ToyModelBackgroundWrapping} is exact. The pull-back to the world-volume of its three-form potential $C$ can be written as 
\begin{align}
    C_3 &= \omega R \bar{R}^2 C^{H}_3 + \mathcal(\omega^0) 
\end{align}
with $C^{H}_3 = - \frac{1}{3!} \omega R \bar{R}^2 F_{klmN}(X^n) \ \sigma^N \mathrm{d}X^k \wedge \mathrm{d}X^l \wedge \mathrm{d}X^m$. Note that the leading order does not depend on longitudinal fluctuations $X^M(\sigma)$.

The membrane action \eqref{eq:MembraneAction} at leading orders becomes
\begin{equation}
    S = \left( S_H[X^k] + \frac{R^2}{\bar{R}^2} S_\tau[X^K] + \mathcal{O} \left( \frac{1}{\omega},\frac{\bar{R^2}}{R^2 \omega^2} \right) \right) \label{eq:NGActionLimit}
\end{equation}
in the limits $\omega \rightarrow \infty$ and $\frac{\bar{R}^2}{R^2 \omega^2} \rightarrow 0$, with
\begin{align*}
    S_H &= T_0 \int \mathrm{d}^3 \sigma \left( \frac{1}{2} H_{mn}(X^k) \partial_M X^m \partial_N X^n \ \eta^{MN} \right. \\
    &{} \qquad \qquad \left. - \frac{1}{3!} F_{klmN}(X^n) \ \sigma^N \partial_K X^k \partial_L X^l \partial_M X^m \epsilon^{klm} \right), \\
    S_\tau &=  T_0  \int \mathrm{d}^3\sigma  \left( \partial_K X_L \partial^K X^L + (\partial_K X^K)^2 \right).
\end{align*}
The first two terms in \eqref{eq:InducedMetricDetExpansion} account for total derivatives in the action and will be neglected because they do not contribute to the equations of motion. In order to obtain a finite action, the following scaling of the membrane tension in terms of a finite $T_0$ was introduced:
\begin{equation}
    T \sim \frac{T_0}{\bar{R}^2 R \omega}.
\end{equation}
$S_H$ depends only on the transverse and $S_\tau$ only on longitudinal fluctuations. Hence, as striven for, at leading orders we remain with models with trivial world-volume gravity (the world-volume metric becomes $\eta_{MN}$) and decoupling of transversal and longitudinal dynamics. 

\paragraph{The fate of reparameterisation invariance.} At the level of the original membrane action, the model possesses reparameterisation invariance. In particular, one might be tempted to choose \textit{static gauge}, $X^M \equiv 0$, for the longitudinal directions. In general, the theory after the limit \eqref{eq:NGActionLimit}, does not show reparameterisation invariance. Moreover, $S_\tau$ depends explicitly on longitudinal fluctations. It is not surprising that the resulting theories after the singular 'flat' limit $\gamma \rightarrow \eta$ (at leading order) do not exhibit reparameterisation invariance and depend on the choice of parameterisation from where we start the limit. 

Nevertheless, if one takes the scaling of $R$ and $\bar{R}$ as
\begin{equation}
    \frac{R}{\bar{R}} \rightarrow 0,
\end{equation}
interestingly, the leading order dynamics in \eqref{eq:NGActionLimit} does \textit{not} depend on the longitudinal fluctations, and hence this case is compatible with the choice of static gauge for the original membrane.

\paragraph{Interpretation of the limit.} There are several ways to choose interesting concrete realisations and interpret the combined scaling limits:
\begin{equation}
    \omega \rightarrow \infty, \quad  \frac{\bar{R}}{R \omega} \rightarrow 0, \quad T \sim \frac{T_0}{\bar{R}^2 R \omega} .\label{eq:Rscaling}
\end{equation}
Let us outline one interpretation here. One can solve \eqref{eq:Rscaling} with the non-Lorentzian parameterisation of the target space introduced in appendix \ref{app:NR}. For this, let us introduce the dimensionless scale $c$ with
	\begin{equation}
			\omega = c^\alpha , \quad  R^2 = c^2, \quad \bar{R}^2 = \frac{1}{c}.
	\end{equation}
	In order to satisfy \eqref{eq:Rscaling}, the scaling of $\omega \sim c^\alpha \rightarrow \infty$ can be chosen accordingly in terms of $c$. There are two choices
	\begin{itemize}
		\item a non-relativistic limit $c \rightarrow \infty$: $\alpha > 0$. 

		\item a hyper-relativistic (Carollian) limit $c \rightarrow 0$: 
        
        $\alpha < - 3/2$. Remarkably, this case leads to $\frac{R}{\bar{R}} \rightarrow 0$ and, therefore, to the total absence of the longitudinal fluctuations in \eqref{eq:NGActionLimit}.
	\end{itemize}
Interestingly, in both cases this corresponds to a \textit{large energy} ($E \sim \omega R \rightarrow \infty$) and \textit{small tension} ($T \sim \frac{T_0}{\omega} \rightarrow 0$) limit. Nevertheless, many other ways of interpreting \eqref{eq:Rscaling}, with other physical consequences, are possible.

\section{The MZW-model as decoupling limit of the membrane}
\label{chap:MZWfromMembrane}

\subsection{A toy model background}
In order to obtain the MZW-model from the setting in the previous section, we specify $\mathcal{M}_H$ to be a Lie group $G$, in particular we choose:
\begin{align}
    \mathrm{d}s^2 &= R^2 \eta_{MN} \mathrm{d}\hat{X}^M \mathrm{d}\hat{X}^N + \bar{R}^2 \kappa_{ab} e^a e^b , \label{eq:ToyModelBackgroundMZW} \\
     F &= - R \bar{R}^2 \frac{1}{3!} f_{abc} e^a \wedge e^b \wedge e^c \wedge v .  \nonumber
\end{align}
$e^a = {e_m}^a \mathrm{d}\hat{X}^m$ are the Maurer-Cartan one-forms on $G$, $f_{abc}$ are the structure constants for the Lie algebra $\mathfrak{g}$ of $G$, and $\kappa$ is the Killing form on $\mathfrak{g}$. $v$ is a one-form on $\mathcal{M}_\tau$. We choose $v = \mathrm{d}\hat{X}^y$ for $y = 0,1$ or $2$. This choice will be related to the choice of $v$ in the MZW-model \eqref{eq:MZW}. In particular, one is left with
\begin{equation}
    \int C_3 = \frac{1}{3}\omega R \bar{R}^2 \int \sigma^y \text{tr} \left( j \wedge [j \ \overset{\wedge}{,} \ j ] \right) + \mathcal{O}(\omega^0)
\end{equation}
on the world-volume after applying \eqref{eq:ExpansionWrapping}. 

One can easily see that for this particular background the $S_H$, the part of the membrane action \eqref{eq:NGActionLimit} containing the transverse fluctuations, becomes the MZW model action 
\begin{equation}
    S_H = S_{MZW}.
\end{equation}
In appendix \ref{app:Polyakov}, we present how a similar statement can be obtained for the Polyakov formulation of the membrane at the level of the equations of motion.

\subsection{Embedding into a supergravity solution}
The toy model background \eqref{eq:ToyModelBackgroundMZW} is not a solution to eleven-dimensional supergravity, but can be embedded in one. The prototypical example will be provided by the eleven-dimensional uplift of the AdS$_3 \times$S$^3 \times$T$^4$ background, due to the presence of a group manifold being supported by four-form flux:
\begin{align*}
    \mathrm{d}s^2 &= \bar{R}^2 \left( \mathrm{d}s^2_{\text{AdS}_3} + \mathrm{d}s^2_{\text{S}^3} \right) + \mathrm{d}s^2_{T^4} + \mathrm{d}z^2 \\
    F &= 2\bar{R}^2 \left( Vol(\text{AdS}_3) + Vol(\text{S}^3 ) \right) \wedge \mathrm{d}z
\end{align*}
where $z$ is the coordinate of the M-theory circle. Moreover, let us single out one of the directions of the $T^4$, denoted by $y$, and introduce the scales $R$ and $\bar{R}$
\begin{align}
    \mathrm{d}s^2 &= \bar{R}^2 \left( \mathrm{d}s^2_{\text{AdS}_3} + \mathrm{d}s^2_{\text{S}^3} \right) + R^2 \left( \mathrm{d}y^2 + \mathrm{d}z^2 \right) +  \mathrm{d}s^2_{T^3} \nonumber  \\
    F &= 2 R \bar{R}^2 \left( Vol(\text{AdS}_3) + Vol(\text{S}^3 ) \right) \wedge \mathrm{d}z . \label{eq:AdS3S3uplift}
\end{align}
With the following expansion (in a Poincar\'e patch of AdS$_3$) for example:
\begin{align}
    \hat{X}^\mu(\sigma) &= \left( X_{AdS_3}(\sigma) ; X_{S^3}^m(\sigma) ; X_y(\sigma) , X_z(\sigma) , X_{T_3}(\sigma) \right) \nonumber \\
    X_{AdS_3}(\sigma) &= (t(\sigma),x(\sigma),r(\sigma)) = \left( \omega \frac{R}{\bar{R}} \sigma^0 , 0 , 0 \right) \label{eq:ExpansionAdS3S3} \\
    X_y (\sigma) &= \omega \sigma^1, \quad X_z(\sigma) = \omega \sigma^2, \quad X_{T^3}(\sigma) = const. \nonumber
\end{align}
Hence, the focus lies on the transversal fluctuations $X^m_{S^3}$ around the $S^3$, neglecting the other (finally decoupled) fluctuations -- both along the remaining transversal directions, but also the longitudinal ones. The only significant difference from the toy model case is that the time-like direction lies in AdS$_3$ which scales like the transverse direction. In the above expansion, this is rectified by parameterisation of $t(\sigma)$ including the ratio $\frac{R}{\bar{R}}$. Hence, for a large wrapping limit $\omega \rightarrow \infty$, and suitable limits for $R$ and $\bar{R}$, as discussed above \eqref{eq:Rscaling}, one reproduces the action of the MZW-model for fluctations $X_{S^{3}}^m$.

\section{Outlook}
In this letter, we proposed a construction that shows that a certain regime of the (bosonic) membrane can be described by a classically integrable theory, the $(1+2)$-dimensional MZW-model, which has been studied in some detail in the literature. In this way, this proposal also offers a physical realisation of this model. Some crucial questions for future research will be
\begin{itemize}
    \item What is the physical interpretation of the conserved charges \cite{Ioannidou:1995he} and integrable hierarchy \cite{Dimakis:1999sw,wu2008cauchyproblemwardequation} of the MZW-model for the membrane? Do these also have meaning outside of this regime?
    \item What is the physical interpretation of the solitons \cite{Ward:1990vc,Ward:1995he,Ioannidou:1999} of the MZW-model on the membrane?
    \item Is there a dual holographic picture, corresponding to the expansion \eqref{eq:ExpansionAdS3S3} of the eleven-dimensional uplift of the NSNS AdS$_3 \times$S$^3$ background?
\end{itemize}
A background with very similar features to \eqref{eq:ToyModelBackgroundMZW} would actually exist as a solution in type IIa supergravity: the T-dual to the pure RR IIb AdS$_3 \times$S$S^3 \times T^4$ background. The D2-brane DBI action in this background should possess a similar limit to the one dicussed in section \ref{chap:MZWfromMembrane} for the membrane.

For integrable two-dimensional $\sigma$-models, generalised geometry resp. O$(d,d)$-duality was used in order to construct and understand new integrable models \cite{Osten:2019ayq,Borsato:2021gma,Borsato:2021vfy}. For the membrane a similar formulation, geometrising an underlying $E_{d(d)}$-structure, has been obtained and discussed in the literature \cite{Hatsuda:2012vm,Sakatani:2020iad,Sakatani:2020umt,Osten:2021fil,Hatsuda:2023dwx,Osten:2024mjt}. It might be fruitful to systematically study non-Lorentzian expansions of this setting in order to obtain new regimes of membrane theory that are described by integrable $(1+2)$-dimensional field theories. In general, the advent of novel higher-dimensional integrable theories via decoupling limits could be an interesting consequence of the present work. As mentioned above, another potential path to new higher-dimensional integrable $\sigma$-models could be possible via the recently suggested higher Chern-Simons theory \cite{Schenkel:2024dcd}.

\subsection*{Acknowledgements}
The research of D.O. is part of project No. 2024/55/D/ST2/01205 funded by the National Science Centre (NCN) of Poland. The author thanks Arkady Tseytlin for comments on the first version, the Institute for Theoretical Physics of ETH Zurich for their hospitality when finalising this work, the participants of the 'Workshop on Higher-d Integrability' in Favignana for fruitful discussions about the draft, and the COST action CA22113 for support under the Short-Term Scientific Mission grant E-COST-GRANT-CA22113-2a51a346. Special credits belong to to the anonymous referees of \textit{Physical Review Letters} for their crucial constructive comments on the original manuscript. 

\appendix
\section{The non-Lorentzian expansion of supergravity} \label{app:NR}
The non-Lorentzian nature of the MZW-model suggests the use of non-Lorentzian geometry. In particular, we are interested in the application to membranes in eleven-dimensional supergravity. A finite non-relativistic limit of eleven-dimensional supergravity has been obtained in \cite{Blair:2021waq,Bergshoeff:2024nin} and gives rise to the following expansions of the metric $g$ and three-form gauge field $C$ as series of some dimensionless parameter:
\begin{align}
   g_{\mu\nu} &= c^2 \tau_{\mu}{}^A \tau_\nu{}^B \eta_{AB} + c^{-1} H_{\mu \nu}  ,\label{eq:ExpansionG} \\
   C_{\mu_1 \mu_2 \mu_3} &= c^3 \tau_{\mu_1}{}^A \tau_{\mu_2}{}^B\tau_{\mu_3}{}^C \epsilon_{ABC} + C_{\mu_1 \mu_2 \mu_3}. \label{eq:ExpansionC}
\end{align}
For this expansion, only the non-relativistic limit $c \rightarrow \infty$ was considered in the literature. We take this as a heuristic motivation that an ultra-relativistic or Carollian limit ($c \rightarrow 0$) should take a similar form, motivated by similar considerations in general relavitiy \cite{Hansen:2021fxi}. Both non- and ultra-relativistic limits are of interest in main text. 

$\eta_{AB}$ and $\epsilon_{ABC}$ are the $(1+2)$-dimensional Minkowski metric and totally antisymmetric tensor. The geometry, locally, describes a split into a $(1+2)$-dimensional relativistic and an eight-dimensional non-relativistic space, by three \textit{longitudinal} (so-called clock) one-forms $\tau^A$ and a \textit{transverse} metric $H$. In inverse of these objects is defined by the completeness and orthogonality conditions
\begin{align*}
    H^{\mu \rho} H_{\rho \nu} + {\tau^\mu}_A \tau_\nu{}^A &= \delta_\nu^\mu, \\
    H^{\mu \nu} {\tau_\nu}^A = 0 &= H_{\mu \nu} {\tau^\nu}_A, \quad {\tau^\mu}_A \tau_\mu{}^B = \delta_A^B .
\end{align*}
The relevant examples in this letter are very simplistic, allowing a global split into longitudinal and transverse space with coordinates $\hat{X}^\mu = (\hat{X}^M , \hat{X}^m)$. Moreover, $\mathrm{d} \tau^A = 0$, the condition for the so-called Newton-Cartan torsionless membrane geometry, will always be satisfied here.

\section{The MZW-model from the Polyakov formulation}
\label{app:Polyakov}
The standard action \cite{Bergshoeff:1987cm} of the bosonic membrane in the Polyakov form including a dynamical world-volume metric $\gamma$ is 
\begin{align}
    S &= T \int \mathrm{d}^3 \sigma \left( \frac{1}{2} \sqrt{- \gamma}  \gamma^{MN} \partial_M \hat{X}^\mu \partial_N \hat{X}^\nu g_{\mu\nu} \right. \label{eq:MembraneActionPolyakov} \\
    &{} \qquad + \left. \frac{1}{3!} \epsilon^{KLM} \partial_K \hat{X}^\mu \partial_L \hat{X}^\nu \partial_M \hat{X}^\rho C_{\mu \nu \rho} - \frac{1}{2} \sqrt{-\gamma} \right) \nonumber
\end{align}
with tension $T$ and in a bosonic supergravity background described by a metric $g$ and three-form gauge field $C$. $\sigma^M$ are the coordinates on the membrane world-volume, $\hat{X}^\mu(\sigma)$ the embedding coordinate fields, and $\gamma_{MN}$ is the (dynamical) world-volume metric. The cosmological constant term, as usual, ensures that the world-volume metric becomes the induced metric:
\begin{equation}
    \gamma_{MN} = \partial_M \hat{X}^\mu \partial_N \hat{X}^\nu g_{\mu \nu}.
\end{equation}
In contrast to the string, where reparameterisations and Weyl invariance can be used to gauge-fix the world-volume metric to be flat, this is not the case for the membrane. The equations of motion for the world-volume coordinate fields $\hat{X}^\mu(\sigma)$ are
\begin{align}
    0& = \partial_M (\sqrt{- \gamma}  \gamma^{MN} \partial_N \hat{X}^\mu) + \sqrt{- \gamma}  \gamma^{MN} \Gamma^{\mu}{}_{\kappa \lambda} \partial_M \hat{X}^\kappa \partial_N \hat{X}^\lambda, \nonumber \\
    &{} \quad + \epsilon^{NKL} \frac{1}{3!} {F^\mu}_{\nu \kappa \lambda} \partial_N \hat{X}^\nu \partial_K \hat{X}^\kappa \partial_L \hat{X}^\lambda, \label{eq:EquationsOfMotion}
\end{align}
where $\Gamma$ denote Christoffel symbols and $F= \mathrm{d}C$. Due to the non-trivial nature of the world-volume metric $\gamma$, these equations are quite intractable, in general. After restricting to the general toy model background \eqref{eq:ToyModelBackgroundWrapping} and wrapping ansatz \eqref{eq:ExpansionWrapping} for $\omega \rightarrow \infty$ and $\frac{\bar{R}^2}{R^2 \omega^2} \rightarrow 0$ the induced metric becomes \eqref{eq:InducedMetricExpansion}. 

Including this into the equations of motion \eqref{eq:EquationsOfMotion} for $X^M$ $X^m$, one obtains at leading order 
\begin{align}
   0 &= \omega R \left( 0 + \mathcal{O}\left(\frac{1}{\omega},\frac{\bar{R}^2}{R^2 \omega} \right) \right) \label{eq:EomDecouplingWrapping}  \\
   0 &=  \omega R \left( \partial_M \partial^M X^m + \Gamma_{(H)}^{m}{}_{kl} \partial_M X^k \partial_N X^l \right. \label{eq:EomDecouplingWrapping2} \\
 &{} \quad \left. + \frac{1}{2}  H^{mn} F_{klnN} \partial_K X^k \partial_L X^l \epsilon^{KLN}  + \mathcal{O}\left( \frac{1}{\omega},\frac{\bar{R}^2}{R^2 \omega} \right) \right) . \nonumber
\end{align}
If one specifies to the special background of the form \eqref{eq:ToyModelBackgroundMZW} ($\mathcal{M}_H$ being a Lie group), at leading order the membrane equations of motion, now expressed in terms of currents $j^a = {e_m}^a \mathrm{d} X^m$ become the equations of motion of the MZW model \eqref{eq:MZW}.

Remarkably, at leading order the longitudinal equations of motion vanish identically -- at higher order one would expect additional coupling of longitudinal and transversal fluctuations. But we should note that in this formulation the relative and absolute scaling of both equations is not obvious -- as compared to the formulation in terms of the action \eqref{eq:NGActionLimit}.

As argued in the main text, the limit does not seem to commute with classical reformulations (such as reparameterisations), and also seemingly the formulation as Polyakov or Nambu-Goto type model. Nevertheless, the defining feature of this limit (reproducing the MZW model) is present in any case.

{\small
\bibliographystyle{jhep}
\bibliography{References}}
\end{document}